\documentclass[useAMS,usenatbib,usegraphicx]{mn2e}
\newcommand{\arcdeg}{\degr}
\newcommand{\masyr}{\mbox{mas yr$^{-1}$}}
\newcommand{\muas}{\mbox{$\mu$as}}
\newcommand{\muasyr}{\mbox{$\mu$as yr$^{-1}$}}

\newcommand{\kms}{\mbox{km s$^{-1}$}}
\newcommand{\Jb}{\mbox{Jy bm$^{-1}$}}

\newcommand{\vsig}{\mbox{$\sigma_{v1{\rm D}}$}}

\newcommand{\lesssim}{\mbox{\raisebox{-0.3em}{$\stackrel{\textstyle <}{\sim}$}}}

\newcommand{\Ra}[4]{\mbox{${#1}^{\rm h} \; {#2}^{\rm m} \; {#3}\fs{#4} $}}
\newcommand{\dec}[4]{\mbox{${#1}\arcdeg \; {#2}\arcmin \; {#3}\farcs{#4} $}}

\newcommand{\tablenotemark}[1]{$^{\rm #1}$}
\newcommand{\tablenotetext}[2]{\noindent$^{\rm #1}$ #2}

\newcommand{\PSRJ}{PSR J0205+6449}
\newcommand{\JREF}{J0209+6437}
\newcommand{\JCHECK}{J0228+6721}

\title{The Proper Motion of \PSRJ\ in 3C 58}

\author[Bietenholz et al.]{M.\ F.\ Bietenholz$^{1,2}$, 
V. Kondratiev$^{3,4}$, S. Ransom$^5$, P. Slane$^6$,
N.\ Bartel$^2$, and
S. Buchner$^1$ \\
$^1$Hartebeesthoek Radio Observatory, PO Box 443, Krugersdrop, 1740, South Africa \\
$^2$Department of Physics and Astronomy, York University, Toronto, M3J 1P3, Ontario, Canada \\
$^3$ASTRON, the Netherlands Institute for Radio Astronomy, Postbus 2, 7990 AA Dwingeloo, The Netherlands\\
$^4$Astro Space Center of the Lebedev Physical Institute, Profsoyuznaya str. 84/32, Moscow 117997, Russia\\
$^5$National Radio Astronomy Observatory, Charlottesville, VA 22903, USA \\
$^6$Harvard-Smithsonian Center for Astrophysics, Theory Division, 60
Garden Street, Cambridge, MA 02138, US \\}

\begin{document}

\date{Accepted for publication on MNRAS; \today}

\pagerange{\pageref{firstpage}--\pageref{lastpage}} \pubyear{2011}
 
\maketitle

\label{firstpage}

\begin{abstract}

We report on sensitive phase-referenced and gated 1.4-GHz VLBI radio
observations of the pulsar \PSRJ\ in the young pulsar-wind nebula
3C~58, made in 2007 and 2010\@.  We employed a novel technique where the
$\sim$105-m Green Bank telescope is used simultaneously to obtain
single-dish data used to determine the pulsar's period as well as to
obtain the VLBI data, allowing the VLBI correlation to be gated
synchronously with the pulse to increase the signal-to-noise. The high
timing noise of this young pulsar precludes the determination of the
proper motion from the pulsar timing. We derive the position of the
pulsar accurate at the milliarcsecond level, which is consistent with
a re-determined position from the {\em Chandra} X-ray observations.
We reject the original tentative optical identification of the pulsar
by \citet{ShearerN2008}, but rather identify a different optical
counterpart on their images, with $R$-band magnitude $\sim$24\@.  We
also determine an accurate proper motion for PSR J0205+6449 of ($2.3
\pm 0.3$)~\masyr, corresponding to a projected velocity of only $(35
\pm 6)$~\kms\ for a distance of 3.2~kpc, at p.a.\ $-38$\arcdeg.  This
projected velocity is quite low compared to the velocity dispersion of
known pulsars of $\sim$200~\kms.  Our measured proper motion does not
suggest any particular kinematic age for the pulsar.
\end{abstract}


\section{Introduction}

The supernova remnant 3C~58 (G130.7+3.1) has a filled centre
morphology and a flat radio spectrum, and was classified on this basis
as being pulsar wind nebula \citep[PWN, also plerion;
  e.g.,][]{WeilerP1978} long before a pulsar was seen.  The pulsar,
\PSRJ, was subsequently detected, first in the X-ray
\citep{Murray+2002} and then in the radio \citep{Camilo+2002}. It is
also one of the approximately 100 pulsars that shows pulsed gamma-ray
emission \citep{Abdo+2009}.

\PSRJ\ and 3C~58 are at a distance, $D$, of $\sim$3.2~kpc
\citep{Roberts+1993}\footnote{Note that the distance is somewhat
uncertain, and somewhat larger value is possible.  The pulsar
dispersion measure is approximately twice what is expected at the
distance of 3.2~kpc \citep{Camilo+2002}.}.  
They have traditionally been associated with the historical supernova
of 1181~AD \citep[SN~1181;][]{ClarkS1977, StephensonG1999}, giving
them an age of 830~yr and making them one of the youngest known
supernova remnants and pulsars.  However, recent work has suggested
that 3C~58 is likely considerably older.  In particular, the measured
expansion speeds of both the synchrotron bubble \citep{3C58-2001,
3C58-2006} and of the thermal filaments \citep{Fesen1983, FesenKB1988,
vdBergh1990, RudieF2007}, seem to be considerably lower than expected
for an age of $\sim 830$~yr, suggesting an age of several thousand
years, which larger age is also suggested several other arguments
\citep[see, e.g.,][]{3C58-2006, Chevalier2005}.

\PSRJ\ has a spin frequency of $\sim$15.2~Hz, with a derivative of
$\sim -4.5 \times 10^{-11}$~Hz~s$^{-1}$ \citep{Murray+2002,
Camilo+2002, Livingstone+2009}, so
the characteristic age of the pulsar is 5400~yr.  (We note that if the
pulsar had an initial spin frequency of $\sim$16.3~Hz, the present
spin frequency and spindown rate could be reconciled with an age of
830~yr; however as mentioned, numerous other arguments independent of
the characteristic age suggest a considerably larger age).
It has a high level of timing noise, having exhibited two spin-up
glitches with fractional magnitudes of $\Delta\nu / \nu = 3.4 \times
10^{-7}$ and $3.8\ \times 10^{-6}$ between MJDs 52276 and 53063
\citep{Livingstone+2009}.
The pulse width is $\sim$2.5~ms \citep[at 1.4 GHz,][]{Camilo+2002}.
Although as mentioned, there is some controversy over the age of
3C~58, \PSRJ\ is nonetheless one of the youngest known
pulsars\footnote{Several pulsars or PWNe, including the Crab Nebula,
  G21.5$-$0.9 \citep{G21.5expand} and Kes~75 \citep{Gotthelf+2000} are
  thought to be only around 1000~yr old, but the vast majority of
  known pulsars have ages $>$10,000~yr.}.

\PSRJ\ is a particularly interesting case because of the presence of
the easily observable pulsar wind nebula, 3C~58\@. Such nebulae, seen
for only a handful of pulsars, provide important diagnostics for young
pulsars, giving insight into the winds which carry away the pulsar
spindown energies.

Knowing \PSRJ's proper motion is important for determining its exact
birthplace as well as the nature of its interaction with its
surrounding PWN\@.  More generally, the origin of the high space
velocities of pulsars is an interesting question, and therefore
knowledge of \PSRJ's proper motion is particularly important because
of its young age.

\section{Pulsar Timing with GBT: Observations and Results}

We obtained two sessions of pulsar timing observations at the
$\sim$105-m diameter Green Bank Robert C. Byrd telescope at Green Bank
(GBT), which were carried out simultaneously with the VLBI sessions.
We used a new observing technique where the data from the GBT was
simultaneously used for both pulsar timing and the VLBI observations.
The VLBI data are described in the next session.

On 2007 Apr 25, we observed \PSRJ\ with the GBT Spigot back-end
\citep{kel+05} for a duration of 10.8~hrs.  The Spigot measures
3-level autocorrelations from each of 2 polarisations for 1024 lags
and integrates them for 81.92~$\mu$s before summing them and writing
them to disk.  The total bandwidth covered 800~MHz, centred at
1650~MHz, approximately 500~MHz of which (from $\sim 1350 - 1850$~MHz)
were free enough from interference to be used for the timing analysis.
We let the Spigot continue to take data while the GBT occasionally
moved to and from a calibrator source as part of the VLBI scheduling
(see below).  For these times, which corresponded to about 35\% of the
total duration, we simply zero-weighted the resulting data from the
pulsar.  A standard pulsar timing analysis using {\tt TEMPO} provided
us with a pulsar spin ephemeris which we used for pulsar gating of the
interferometric data during correlation.  We measured a barycentric
pulsar spin period of 0.0657176116(4)~s at epoch MJD 54215.3220 (UTC).

We performed a very similar observation with the newer pulsar back-end
GUPPI \citep{rdf+09} on 2010 Oct 18, where we observed \PSRJ\ for
8.9~hrs.  GUPPI recorded the 8-bit, summed polarisations from 2048
channels every 192~$\mu$s covering 800~MHz of bandwidth (of which the
highest $\sim$600~MHz was usable).  After zero-weighting data when the
GBT was off of the pulsar's position, a timing analysis determined the
barycentric pulsar spin period to be 0.0657388555(6)~s at epoch MJD
55487.8302 (UTC).

For our timing analyses we used a simple two-Gaussian model as our
template profile where the peak of the pulsar flux occurred very near
to spin phase 0.5\@.  We assumed a pulsar position, dispersion measure,
and instantaneous spin-down rates consistent with those measured by
\citet{lrc+09}.

\begin{figure}
\centering
\includegraphics[width=0.8\linewidth]{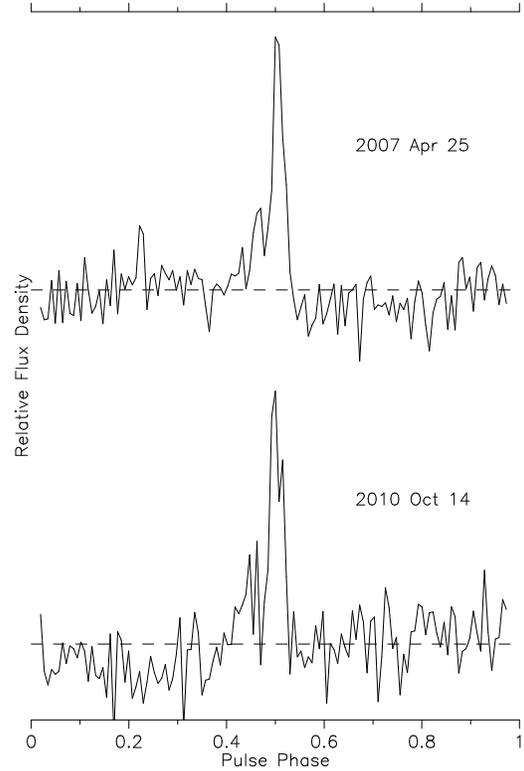}
\caption{Pulse profiles obtained with GBT at our two observing epochs.
The profiles have 128 bins, and have been shifted to so that the peaks
occur at phase = 0.5, and the zero-level in each profile is indicated
by the dashed horizontal line.  We do not consider the differences
between the two profiles significant.}
\label{fgbtprofile}
\end{figure}

We show the average pulse profiles obtained in the two observing
sessions in Figure~\ref{fgbtprofile}.  As the pulsar is quite weak,
the signal-to-noise ratio is relatively low, and consequently we do
not consider the differences between the two profiles significant.
The FWHM duty-cycle the pulse is $\sim$6\%.  

\section{VLBI Radio Observations}
\label{svlbi}

We obtained two epochs of 1.4-GHz VLBI observations
of \PSRJ\ using the ``High Sensitivity Array'', which consisted
of the NRAO
Very Long Baseline Array (VLBA) augmented by the GBT ($\sim$105~m
diameter) and Effelsberg (100~m diameter) telescopes.  For the first
session, on 2007 April 25 (program code BB241), we obtained 10~hours
of VLBI data, while for the second session on 2010 October 18 (program
code BB295), we obtained 12 hours.  In each session we recorded both
senses of circular polarisation, and we used 2-bit sampling at a bit
rate of 512~Mbit~s$^{-1}$, for an effective bandwidth per polarisation
of 64~MHz.

The data from 2007 were correlated with NRAO's VLBA processor, while
those from 2010 were correlated with the DiFX processor
\citep{Deller+2011a}.  The analysis was carried out with NRAO's
Astronomical Image Processing System (AIPS)\@.  During correlation, we
gated the correlator using the pulsar timing extracted from the GBT
data, and accumulated several sets of VLBI data with different gating:
firstly, the un-gated data, and secondly, the ``on-pulse'' gated data
where the gated bin had a width of 8\% of the pulsar period
(distributed symmetrically about the peak).  In addition, for the 2010
data, correlated with the DiFX correlator, we also accumulated data in
12 different bins evenly distributed in pulse phase, with each bin
therefore having a width of 8.33\% of the period.  In all data sets we
used a correlator integration time of 3.0~s.

We used VCS2 \JREF\ \citep{Fomalont+2003},
0.5\arcdeg\ away from \PSRJ, as our primary phase-reference source,
but we included some observations of an astrometric check source, JVAS
\JCHECK (4C~67.05, ICRF J022850.0+672103; IERS B0224+671), which is a
source in the International Celestial Reference Frame catalogue whose
position is accurately known \citep{FeyGJ2009}, and which is
3.4\arcdeg\ away from \PSRJ.  For the phase-centre position for the
pulsar observations, we use the position of the pulsar given in
\citet{SlaneHM2002}, \Ra{02}{05}{37}{92}, \dec{64}{49}{42}{8}.
Mostly we used a cycle time of 3.8 minutes, with 2.5~minutes being
spent on \PSRJ.  The calibration of the VLBI data was done using
standard procedures using NRAO's AIPS package.  The flux density
calibration was done through measurements of the system temperature at
each telescope, and the antenna amplitude gains were subsequently
improved through self-calibration of the reference sources.

Some sporadic radio-frequency interference (RFI) was seen,
particularly in the frequency range 1617 to 1625 MHz.  We clipped out
visibility points with anomalously high amplitudes, removing $\lesssim
1$\% of the data.  
We also deleted any data taken when either of the antennas were at
elevations of $<10$\arcdeg.
During our 2007 observing run, the GBT failed to observe \JREF\ for
three periods of approximately one hour.  We discarded all the GBT
data for all sources from these periods.

The fringe-fitting and phase calibration was done using the (un-gated)
data from our calibrator sources.  We estimated the ionospheric delay
using the AIPS task TECOR using IONEX data from the Crustal Dynamics
Data Information System archive\footnote{\tt
http://cddis.gsfc.nasa.gov} of the Goddard Space Flight Center.
The earth orientation parameters used during the correlation were
extrapolated.  We corrected to more accurate ones subsequently made
available from the United States Naval Observatory.  The resulting
calibration was then interpolated to the times of the \PSRJ\ as well
as the \JCHECK\ observations, and applied to both the gated and
un-gated visibility data for \PSRJ.

\subsection{Astrometric Reference Sources}

\begin{figure}[t]
\centering
\includegraphics[width=\linewidth]{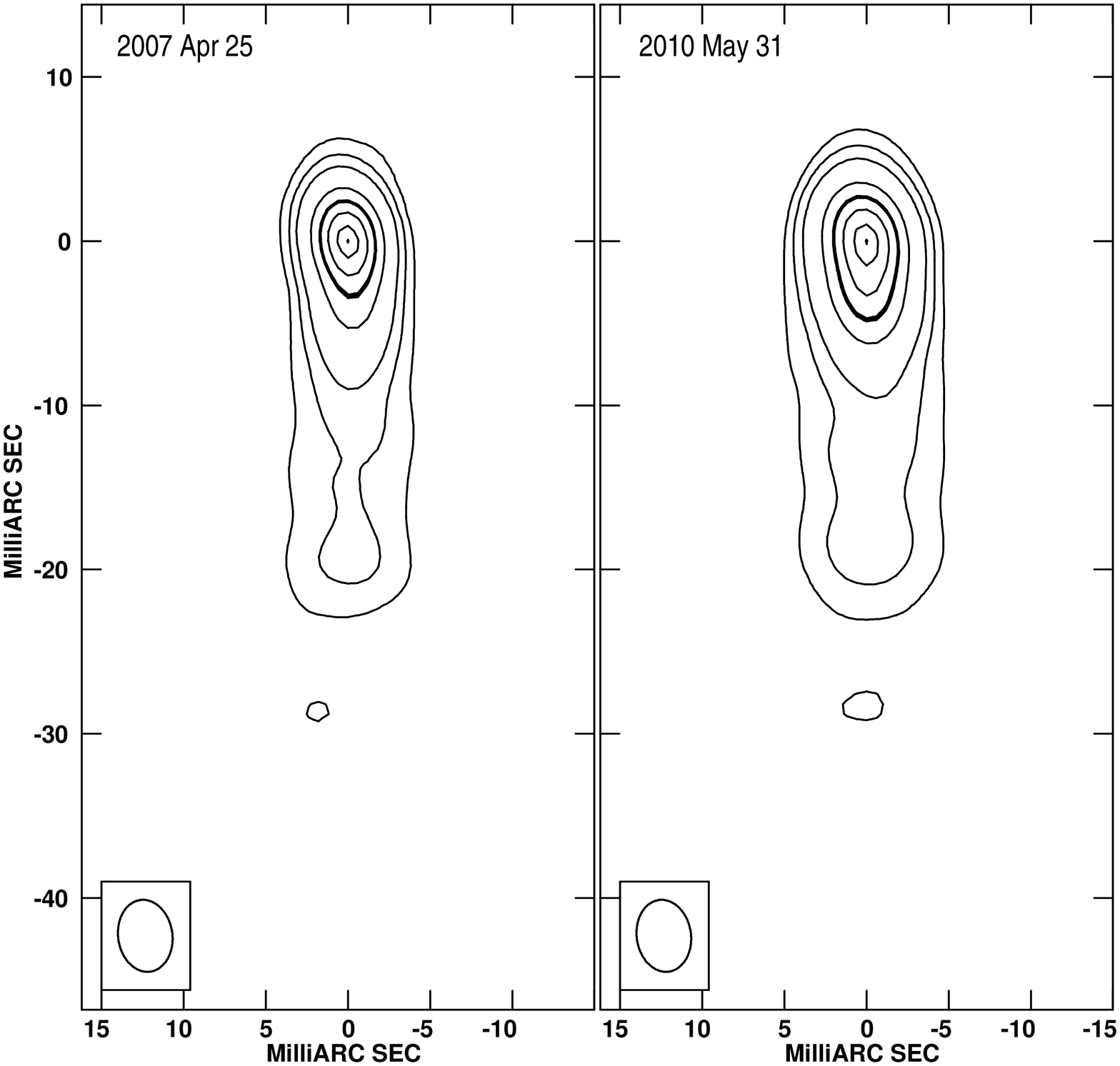} 
\caption{VLBI images of \JREF\ on our two observing epochs.  The
  contours are drawn at 2, 5, 10, 30, {\bf 50}, 70, 90 and 99\% of the
  peak brightness, with the 50\% contour being emphasized.  The two
  images have been convolved with the same restoring beam of FWHM $4.4
  \times 3.3$ mas at p.a.\ 8\arcdeg\ indicated at lower left. The peak
  brightness was 81 and 59 m\Jb\ for 2007 and 2010, respectively.  In
  each panel, the origin is placed at the peak-brightness point, which
  we use as the astrometric reference point.}
\label{fj0209}
\end{figure}

\begin{figure}
\centering
\includegraphics[width=\linewidth]{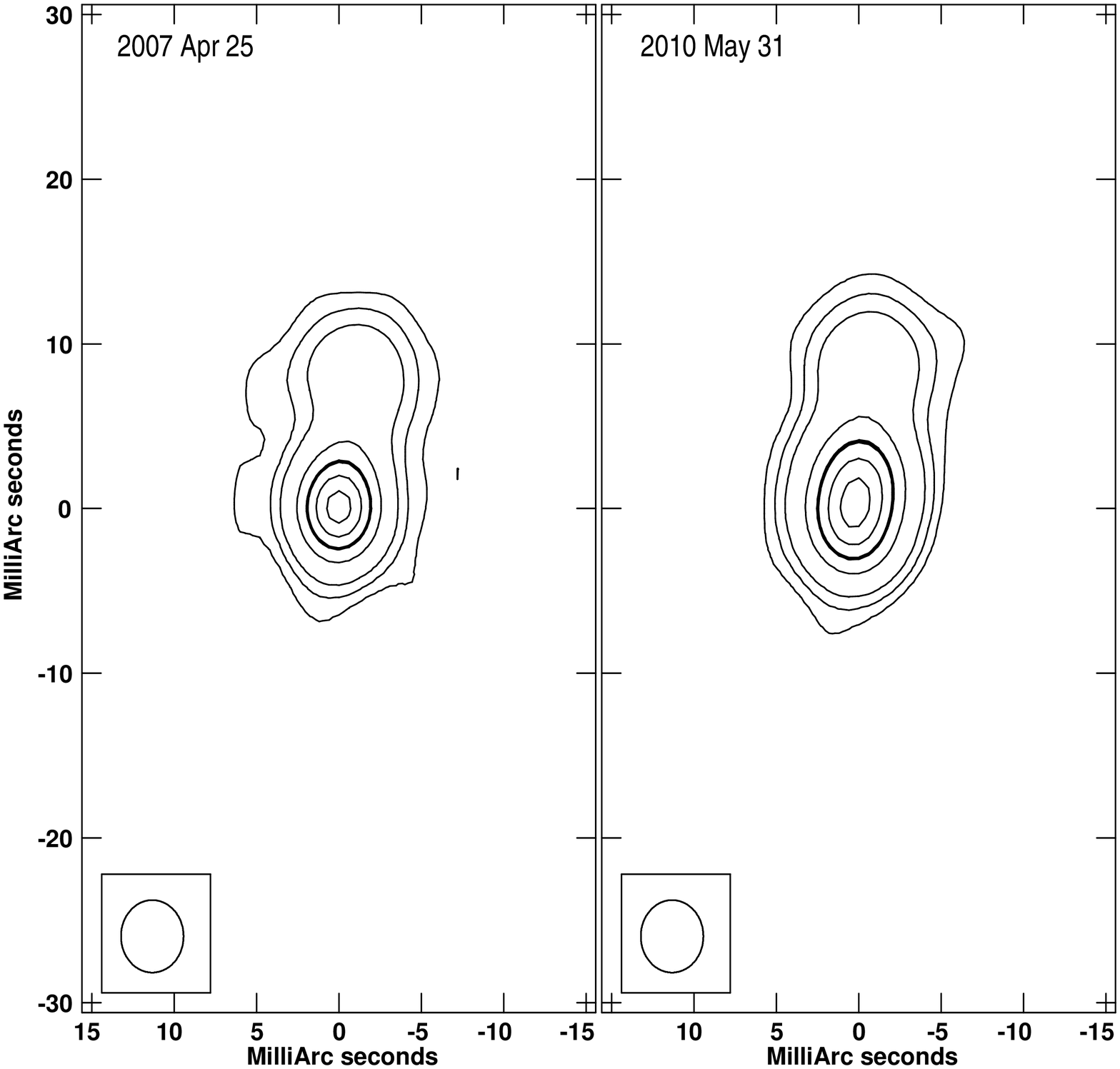} 
\caption{VLBI images of \JCHECK\ on our two observing epochs. The
  contours are drawn at 2, 5, 10, 30, {\bf 50}, 70, 90 and 99\% of the
  peak brightness, with the 50\% contour being emphasized.  The two
  images have been convolved with the same restoring beam of FWHM $4.4
  \times 3.8$ mas at p.a.\ 0\arcdeg\ indicated at lower left. The peak
  brightness was 601 and 421 m\Jb\ for 2007 and 2010, respectively.
  The visibility data were self-calibrated in phase before imaging to
  best show the intrinsic structure.}
\label{fj0228}
\end{figure}

Our primary phase-reference source was \JREF.  We take, as the
astrometric reference point for \JREF, the peak brightness position,
for which we take the coordinates \Ra{02}{09}{35}{98806(20)},
\dec{64}{37}{25}{7701(06)}\ from the VCS2 Survey \citep{Fomalont+2003},
where the numbers in parentheses give the uncertainty in the last two
digits. Unfortunately, this source turned out to be somewhat resolved.
We show the VLBI images of \JREF\ on our two observing epochs in
Figure~\ref{fj0209}.  The source is elongated in the north-south
direction, with the 50\% contour having an extent of approximately
6~mas.

It is possible that the peak brightness position, and therefore our
astrometric reference point, varies with time due to changes in
the source morphology \citep[see, e.g.,][]{M81-2000, GPB-III}

As mentioned, we also included some observations of an astrometric
check source, \JCHECK.  Like our observations of \PSRJ, we
phase-reference our observations of \JCHECK\ to \JREF.  We give the
peak brightness positions of \JCHECK, as well as the ICRF position in
Table~1.

\begin{table*}

\begin{minipage}[t]{0.7\textwidth}
\caption{Position of \JCHECK.}
\begin{tabular}{l c l@{ }l@{ }l l@{ }l@{ }l c c}
\hline
Observations & Frequency& \multicolumn{3}{c}{RA (J2000)}  & \multicolumn{3}{c}{Decl. (J2000)} 
  & \multicolumn{2}{c}{Offset(mas)\tablenotemark{a}}\\
  &(GHz) & hr & m & s & ~\arcdeg  & ~\arcmin & ~\arcsec & RA & decl. \\
\hline
ICRF \citep{FeyGJ2009} & 8.4 & 02 & 28 & 50.05148948 & 67 & 21 & 03.0293039 \\
2007 VLBI obs.\tablenotemark{b} & 1.4 & 02 & 28 & 50.0516247
                                                     & 67 & 21 & 03.033364 & 0.78 & 4.06  \\
2010 VLBI obs.\tablenotemark{b} & 1.4 & 02 & 28 & 50.0516134
                                                     & 67 & 21 & 03.035272 & 0.72 & 5.97 \\
\hline
\end{tabular}
\\
\tablenotemark{a}{Offset of measured position from the ICRF position.}\\
\tablenotemark{b}{Positions measured from peak-brightness point, and measured
relative to \JREF, which is assumed to be at the VCS2 position (see text).}\\
\end{minipage}
\end{table*}

Unfortunately, this source is also extended at the mas level in the
N-S direction \citep{Lister+2009a, Romney+1984, PerleyFJ1980}.
Moreover, this source is known to have a somewhat variable morphology
\citep{Lister+2009c, Bondi+1996, Padrielli+1986}, with both changes in
the size of the central component, and motions of other components
being observed with magnitudes of up to $\sim 0.3$~\masyr.  We show
images of \JCHECK\ in Figure~\ref{fj0228}.

\subsection{\PSRJ}

The pulsar was expected to be quite faint: even in the images made
from gated VLBI data, we expected a signal-to-noise ratio of
$\lesssim$10\@.  Furthermore, the pulsar position is not accurately
known {\em a priori}.  In order not to bias our analysis by searching
only near an assumed position, we therefore chose to image a fairly
large region.
We imaged \PSRJ\ directly from the calibrated multi-channel VLBI data
without any averaging of the visibility data in frequency.  Since our
channel width was 0.5 MHz, the field of view over which bandwidth
smearing is small is $>$10\arcsec, which is adequate for our purposes.
We made images of $4096 \times 4096$ pixels, with the pixels being
$0.8 \times 0.8$~mas, centred on our initial position estimate for
\PSRJ, which was the position was used in the correlation, and 
was the one given by
\citet{SlaneHM2002} of \Ra{02}{05}{37}{92}, \dec{64}{49}{42}{8}
(J2000).  Our images therefore span a region $\pm \sim2\arcsec$ from
this estimated position, which should be larger than any expected
uncertainty in the X-ray derived position.  We used natural weighting
to obtain the highest signal-to-noise ratio.  We evaluated the image
rms brightness over the central 90\% of the image area, excluding a
narrow region around the edge where the rms is typically slightly
higher due to gridding artifacts.

In Table~\ref{tpsrposn} we give the signal-to-noise ratios of the
brightness extrema in the various images.  In the images made from
un-gated visibility data, the positive and negative extrema were of
similar magnitude, being $\leq 5.5\,\sigma$ where $\sigma$ was the
image brightness rms.  In the images made from the on-pulse gated
data, for both epochs, the negative extrema were of a similar
magnitude.  For both epochs however, the images made from the on-pulse
data showed a single positive peak that was $\geq 6.5\,\sigma$, or
larger in magnitude than the negative extremum by $\geq 1\,\sigma$.
The number of independent sky positions sampled by our images can by
estimated by dividing the image area by the FWHM area of the fitted
beam, and was $\sim$650,000\@.  If the pixel brightnesses are
Gaussian-distributed, the likelihood of obtaining a value in excess of
$\pm 6\sigma$ in 650,000 trials is then $\sim$0.1\%, although we
note that the distribution of pixel brightnesses may not be accurately
Gaussian at these large deviations from 0.

Furthermore, the two positive extrema seen in the images made from the
gated data were quite close to one another on the sky, being within
6~mas of each other on the sky.  Since the total area we examined was
$\sim 10^7$~mas$^2$, the odds of random peaks occurring so close
together would be $\sim 80000$ to 1 (which probability is independent
of the actual distribution of pixel brightnesses).

\begin{figure*}
\centering
\includegraphics[width=\linewidth]{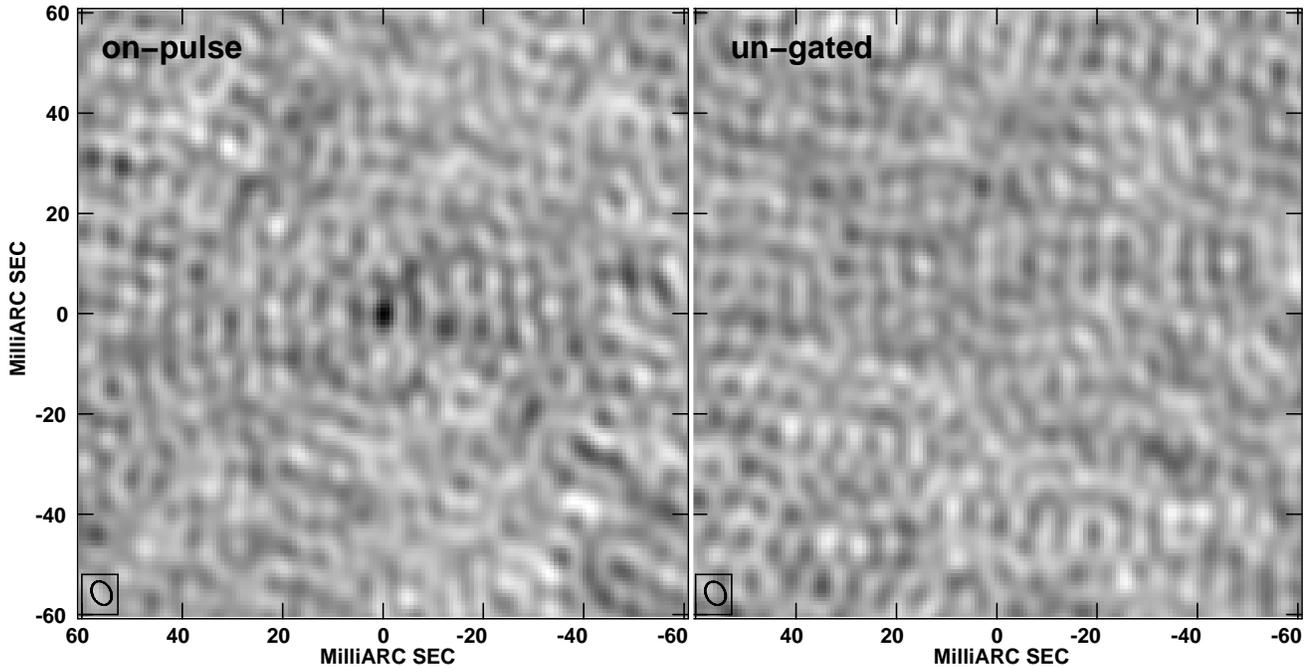} 
\caption{VLBI images of \PSRJ\ from our 2007 observations.  Both
  images are dirty or undeconvolved.  In both panels, the grey-scale
  runs from $-3.8\,\sigma$ (white) to $6.9\,\sigma$ (black), and the
  FWHM of an elliptical Gaussian fit central part of the beam is shown
  at lower left.  The left panel shows the image made from the
  on-pulse gated visibility data, while the right panel shows the
  image made from the un-gated visibility data.  The images are
  centred on \PSRJ, only visible in the left panel, which is at
  \Ra{02}{05}{37}{92322}, \dec{64}{49}{41}{3319}.}
\label{fpsrimage}
\end{figure*}

In Figure~\ref{fpsrimage} we show a section of the images made from
the gated on-pulse and un-gated visibility data sets from 2010\@.
Although the signal-to-noise is not high, the pulsar can be clearly
seen in the on-pulse image, but not in the corresponding un-gated one.

\begin{table*}
\caption{Position of \PSRJ.}
\begin{minipage}[t]{0.7\textwidth}
\begin{tabular}{l c l l l@{ }l@{ }l  l@{ }l@{ }l}
\hline
Observation  & Gating & \multicolumn{2}{c}{Image Extrema}  
 & \multicolumn{3}{c}{RA (J2000)}\tablenotemark{a}
 & \multicolumn{3}{c}{Decl. (J2000)}\tablenotemark{a} \\
Epoch        &       & \multicolumn{2}{c}{(in units of image rms)}  &
  hr & m & s & ~\arcdeg  & ~\arcmin & ~\arcsec  \\
\hline
2007 & Off-pulse & $-5.1$ & +5.1 & \multicolumn{6}{c}{(not detected)}  \\ 
2007 & On-pulse  & $-5.0$ & +6.8 & 02 & 05 & 37.92322 (1) & 64 & 49 & 41.3319 (4) \\ %
2010 & Off-pulse & $-5.5$ & +5.2 & \multicolumn{6}{c}{(not detected)} \\ 
2010 & On-pulse  & $-5.1$ & +6.5 & 02 & 05 & 37.92247 (1) & 64 & 49 & 41.3343 (4) \\

\hline
\end{tabular}
\\
\tablenotetext{a}{The position of the brightness peak of \PSRJ.  The
  digit in parentheses gives the statistical uncertainty in the last
  digit of the corresponding coordinate value; note that the total
  uncertainty is dominated by a systematic component discussed in the
  text.}
\end{minipage}
\label{tpsrposn}
\end{table*}
Despite the modest signal-to-noise ratio, we think that the detections
are firm for several reasons.  Firstly, in each of the ``on-pulse''
images, the positive extremum in the image (excluding a
$\sim$100-pixel strip around the edge where the rms is somewhat higher
due to numerical artifacts) is $\geq 1\,\sigma$ higher than the
corresponding negative one, which is not expected to happen by chance.
Secondly, no source is visible in either of the un-gated images at
this location.  Any background source should be much more strongly
detected in the images made from the un-gated data since the gated
visibility data represents only $\sim$8\% of the total observations.

\begin{figure}
\centering
\includegraphics[width=\linewidth]{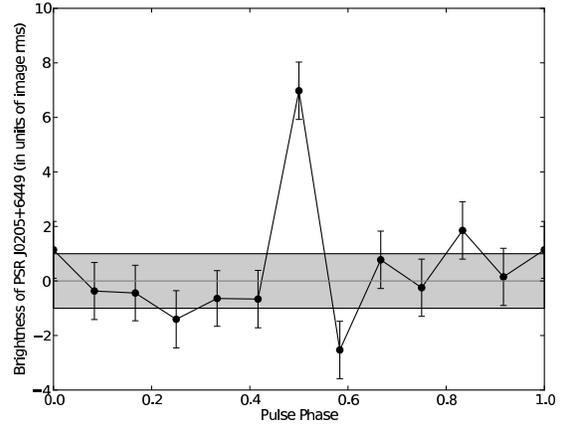} 
\caption{The pulse profile of \PSRJ: the image brightness, given in
  units of the image rms $(\sigma$) at the location of \PSRJ\ in the
  images made from the data in each of 12 bins distributed across the
  pulsar period.  The first bin on the left is duplicated on the right
  for clarity.  The shaded area shows the $\pm 1\,\sigma$ region.
  Only in the bin at pulse phase = 0.5 is a signal substantially in
  excess of the noise visible.}
\label{fpulseprof}
\end{figure}

As an additional check, for the 2010 observations, we imaged
separately the visibility data from the 12 bins spaced evenly across
the pulsar period.  In Figure~\ref{fpulseprof}\ we show brightness at
the location determined from the on-pulse image as a function of the
pulse phase.  The profile can be compared to that obtained from the
GBT pulsar-timing observations shown in Figure~\ref{fgbtprofile}
above.

We note here that position of \PSRJ\ is different than that of the
point source visible in the X-ray as given in \citet{SlaneHM2002} of
\Ra{02}{05}{37}{92}, \dec{64}{49}{42}{8} by $\sim 1\farcs 5$.

\subsection{Proper Motion of \PSRJ}

Using the position determinations in Table~\ref{tpsrposn} we
obtained the displacement of \PSRJ\ between our two observing epochs
to be $-4800 \pm 570,\; +2800 \pm 570$~\muas\ in RA and Decl.,
respectively, where the uncertainties are statistical only.
At a distance of 3.2~kpc, the parallax of \PSRJ\ is expected to be
313~\muas, smaller than the statistical uncertainty, but as the
angular displacement due to parallax is readily calculable we
corrected for it.
The expected shift between our two particular epochs due to parallax
is to be 36 and 499~\muas\ in RA and Decl., respectively.  Correcting
for the parallax, we then arrive at a net displacement of
\PSRJ\ between our two epochs of $-4860 \pm 570,\; +2330 \pm 570$~\muas\
in RA and Decl., respectively (where we have ignored the small
additional uncertainty incurred due to the uncertainty in the
distance).
This displacement corresponds to a proper motion 
of $-1400 \pm 160$~\muasyr\ in RA and $540 \pm 160$~\muasyr\ in
Decl., or $1500 \pm 160$~\muasyr\ at p.a.\ $159 \pm 6$\arcdeg, where
the proper motion is measured relative to the brightness peak of
\JREF, and the uncertainties are statistical only.

Since our reference source, \JREF, is somewhat resolved, we must
carefully assess any possible effect of temporal changes in \JREF\ on
our proper motion measurements. In our VLBI observations, we observed
a check source, \JCHECK\ in a fashion similar to \PSRJ, in particular
similarly phase-referenced to \JREF\@. The positions measured for the
check source were given in Table~1, and suggest a proper motion for
\JCHECK\ of
$-1$ and 550~\muasyr\ in RA and Decl., respectively.  The source
\JCHECK\ is an ICRF source and \citet{Feissel-Vernier2003} finds a
proper motion of $<50$~\muasyr\ at 8.4~GHz over a period of
$\sim$12~years.
However, the lack of secular motions at 8.4~GHz, where the core is
likely more dominant, does not preclude apparent motions at our lower
frequency of 1.4~GHz, where jet components are more likely to dominate
the emission.  Indeed, other authors have reported larger proper
motions for components within \JCHECK: \citet{Bondi+1996} find an
expansion of 300~\muasyr\ while \citet{Lister+2009c} find several
moving components with proper motions of up to 376~\muasyr\ at 14~GHz.
It is therefore possible that our observed proper motion is due to
component motions within \JCHECK.  It is, however, also possible that
our primary reference source, \JREF, is not in fact stable.
Unfortunately, both our reference sources are elongated in the
north-south direction, so that the observed north-south relative
motion between them could be the result of motions along the jet axis
in either source.

As a conservative estimate of $1\,\sigma$ uncertainty on the proper
motion of \PSRJ, as phase-referenced to \JREF, we therefore take the
apparent proper motion of \JCHECK\ found above.  We then arrive at a
final value for the proper motion of \PSRJ\ of $-1400 \pm
160$~\muasyr\ in RA and $540 \pm 575$~\muasyr\ in Decl.  At a distance
of 3.2~kpc, this corresponds to a speed of $23 \pm 6$~\kms.  Note,
however, that this velocity is with respect to the Earth.  We
calculate a more physical value of \PSRJ's projected velocity,
corrected for Galactic rotation and the Sun's motion, in
\S~\ref{sdiscuss} below.

\section{X Ray Astrometry}
\label{sxray}

To determine the X-ray position of \PSRJ, we investigated data from a
deep {\em Chandra} observation carried out between 2003 April 22 and
26 \citep[see][for details and original results from these
  observations]{Slane+2004}. The data from observation IDs 4383, 4382,
and 3832 were reprocessed and cleaned using standard routines from
{\em Ciao} version 4.4\@. The {\tt merge\_all} task was used to create a
merged image from the 317~ks of good exposure time.  The pulsar is
embedded in a bright compact nebula that is slightly asymmetric in the
east-west direction. The centroid of the point source emission is
located at a position of \Ra{02}{05}{37}{93}, \dec{+64}{49}{41.4}. We
identified three X-ray point sources in the field that have
counterparts in the 2MASS catalogue. Comparing the centroid positions
of the X-ray sources with the infrared positions, we find that the
uncertainty in the {\em Chandra} position of the pulsar is
$\sigma_{\rm RA} \sim 0.06$~s and $\sigma_{\rm dec} \sim
0.12\arcsec$. The systematic error associated with contributions from
the compact nebula are estimated to be slightly larger than these
values, but the combined uncertainty is $\lesssim 1$~arcsecond either
direction.

We note that position given above is slightly different from that
determined in the shorter {\em Chandra} observation reported by
\citet{SlaneHM2002}.  We have reprocessed those earlier data
(observation ID 728) as well, and find that the position is in
excellent agreement with the value given above. We conclude that
there were small errors in the initial position reconstructions
reported by \citet{SlaneHM2002}.

\section{Discussion}
\label{sdiscuss}

\begin{figure*}
\centering
\includegraphics[width=\linewidth]{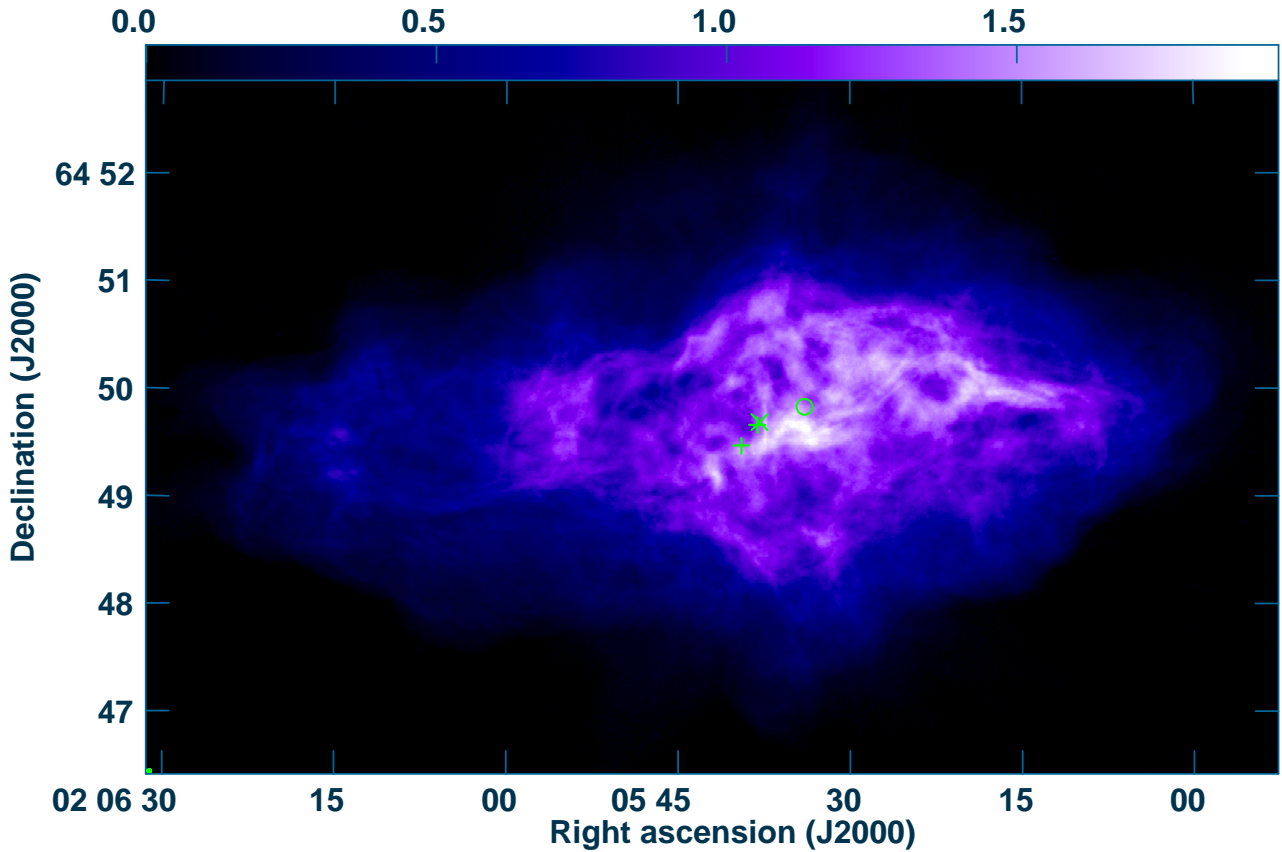}
\caption{A 1.4-GHz VLA radio image of 3C~58, reproduced from
  \citet{3C58-2006}.  The ``$\times$''sign shows the present location
  of \PSRJ, while the two ``+'' signs show the extrapolated position
  at two possible dates of the supernova explosion which gave rise to
  3C~58: at 1181 AD and at 5000 BC, with the one at 1181 AD
  being the one overlapping with the present position ($\times$).  The
  ``$\circ$'' shows the position of the centre of the circular region
  of softer X-ray emission, interpreted as thermal emission from the
  supernova shell, from \citet{GotthelfHN2007}.  The grey-scale is
  labelled in m\Jb, and the FWHM resolution was 1\farcs 4\@. We show
  a detail of the central region in Fig.~\ref{f3c58detail} below.}
\label{f3c58img}
\end{figure*}

Using gated VLBI we imaged \PSRJ\ in the centre of 3C~58.  The pulsar
was detected.  The position of \PSRJ\ we found from the VLBI
observations was $\sim$1\farcs 5 distant from the originally published one
of a compact X-ray source seen in {\em Chandra} ACIS observations
\citep{SlaneHM2002}.
A re-examination of the {\em Chandra} data, however, resulted in an
improved position of the X-ray source which is well within the
uncertainties of that measured with VLBI.

\citet{ShearerN2008} report of deep optical observations of the centre
of 3C~58 using the 4.2~m William Herschel Telescope in La Palma, in
which they detect three unresolved sources with $R$-band magnitudes of
$\sim$24, in addition to the more diffuse synchrotron emission from
the pulsar wind nebula.
\citet{ShearerN2008} tentatively identified their source ``o1'' as
\PSRJ\ on the basis of its coincidence with the published position of
the compact X-ray source.  However, based on our VLBI determination
and the re-determined {\em Chandra} HRC position, it is in fact their object
``o2'', which is almost certainly the optical counterpart of \PSRJ.
The position of o2 is RA = \Ra{02}{05}{37}{93} and Decl. =
\dec{64}{49}{41}{4}, with an uncertainty of $\lesssim 0\farcs 1$,
consistent to within 0\farcs 08 with the pulsar position determined
from VLBI.  \cite{ShearerN2008} give the magnitudes of o2 as
$24.15 \pm 0.07$ in R, $>24.3$ in V and $<25.6$ in B, consistent with
the optical magnitudes estimated for \PSRJ\ based on its spindown
luminosity, distance and the expected Galactic extinction.

\begin{figure}[h]
\centering
\includegraphics[width=\linewidth]{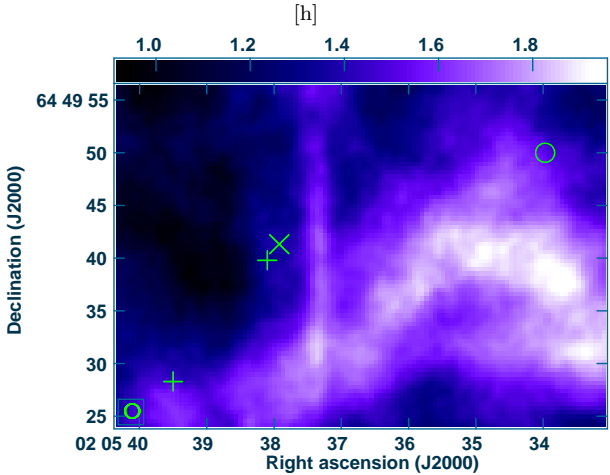}
\caption{Detail of the 1.4 GHz VLA radio image of 3C~58 from
  Figure~\ref{f3c58img} above. The symbols have the same meanings as
  in that figure. The restoring beam FWHM of 1\farcs4 is indicated by
  the circle in the lower left corner.}
\label{f3c58detail}
\end{figure}

In summary, we can regard the detection of \PSRJ\ in the radio,
optical and X-ray bands as firm, with the most accurate position
being that determined from the VLBI observations and given in 
Table~\ref{tpsrposn} above.

In addition to determining the position, we also determined the proper
motion of \PSRJ\ of $-1400 \pm 160$~\muasyr\ in RA and $670 \pm
575$~\muasyr\ in Decl. 
In order to correct this measured value for Galactic rotation and the
Earth's motion with respect to the local standard of rest, we take the
Galactic constants from \citet{Schonrich2012}, namely a flat Galactic
rotation curve with $v = 238$~\kms\ and a distance to the Galactic
centre of 8.27~kpc.  We take also from \citet{Schonrich2012} the Solar
motion of 250~\kms\ in the Galactic plane, solar radial velocity of
14~\kms\ towards the Galactic centre, and a motion of
6.1~\kms\ perpendicular to the Galactic plane.  We take the distance
of \PSRJ\ (from Earth) to be 3.2~kpc.  We can then calculate the
peculiar motion of \PSRJ\ with respect to the standard of rest at its
location as being\footnote{We noted above that the proper motion
  uncertainty is rather larger in Decl. than in RA.  Here we take the
  geometric mean of the two values.} $2.3 \pm 0.3$~\masyr,
corresponding to $35 \pm 6$~\kms\ at p.a.\ $-38$\arcdeg\ (p.a.\ in
equatorial rather than Galactic frame).

This speed is relatively small: Galactic pulsars have a 2-D velocity
dispersion of $200 \sim
300$~\kms\ \citep[e.g.,][]{Hobbs+2005,LyneG1990}.  Even taking the
uncertainty introduced by possible motion of our reference source, the
$3\,\sigma$ upper limit on the proper motion of \PSRJ\ is only
53~\kms.

Can \PSRJ's low tangential velocity be reconciled with the velocity
distribution of other Galactic pulsars?  \citet{Faucher-GiguereK2006}
found that the distribution of space velocities of Galactic pulsars at
birth was consistent with one where each of the three orthogonal
components of the space velocity was distributed exponentially, with
the mean of the absolute value of each component being
$180_{-30}^{+20}$~\kms, and the resulting mean three-dimensional speed
being $380_{-60}^{+40}$~\kms.  Since \PSRJ\ is quite young (we discuss
the age in more detail below), its velocity is probably close to its
birth velocity.  As the multi-dimensional exponential distribution of
\citet{Faucher-GiguereK2006} is difficult to evaluate numerically, we
performed a Monte-Carlo simulation with $n = 10000$ trials, to find
that the chance of finding a pulsar with a tangential velocity as
small as the $35\pm6$~\kms\ measured for \PSRJ\ from such a
distribution is $\sim$2.7\%.

Alternatively, \citet{Hobbs+2005} collected proper motion measurements
for 140 pulsars\footnote{In fact, \citet{Hobbs+2005} collected proper
motion measurements of 175 pulsars, but we exclude the 35 recycled
pulsars from Hobbs' sample as they are much older and have a notably
lower velocity dispersion than the remainder.}.  Of their sample, 8, or
5.7\% had tangential velocities $\leq 35$~\kms.
Finally, some authors have found the distribution of pulsar velocities
to be bi-modal.  For example, \citet{Brisken+2003} find that 20\% of
pulsars form a low-velocity component with a one-dimensional velocity
dispersion, \vsig\ of 99~\kms, while the remainder have a \vsig\ =
294~\kms.  In this case, the probability of a random pulsar having a
tangential velocity as low as 35~\kms\ is 1.8\%.

In conclusion, the small measured proper motion makes \PSRJ\ somewhat
unusual in being amongst the slowest few percent of young pulsars
regardless of whether a bimodal or a unimodal distribution of pulsar
velocities is considered.

Our estimate of the tangential velocity depends on the distance to
\PSRJ\ and 3C~58, which, as mentioned earlier, is somewhat uncertain.
The value of 3.2~kpc of \citet{Roberts+1993} was determined
kinematically from HI absorption.  \citet{Kothes2010} argues for a
distance of $\sim$2~kpc. Adopting a smaller distance would reduce the
tangential velocity estimate by the corresponding factor, and thus
make \PSRJ's low speed even more unlikely.  The low measured angular
speed therefore argues against a distance much lower than 3.2~kpc.

On the other hand, \PSRJ's dispersion measure is approximately twice
that expected for a distance of 3.2~kpc \citep{Camilo+2002},
suggesting perhaps that the true distance is somewhat larger, which
would imply a tangential velocity larger by a factor of ($D/3.2$~kpc).
This question could be resolved with a measurement of \PSRJ's
trigonometric parallax with 10\% accuracy.  Such a measurement is
feasible if an in-beam calibrator can be found \citep[see
e.g.,][]{Chatterjee+2009}, and should be undertaken.

Can we extrapolate back from \PSRJ's present position to determine its
position at the time of the supernova explosion?  To do so requires
knowing the age of \PSRJ.  As noted in the introduction, although it
has traditionally been associated with a supernova in 1181 AD, making
it $\sim$830~yr old, it is probably older, with a likely age of
$\sim$7000~yr \citep{3C58-2006,3C58-2001}, suggesting an epoch of
$\sim$5000~BC for the supernova event.  In Figure~\ref{f3c58img} we show
a radio image of 3C~58, and indicate the present, and extrapolated
epoch 1181 AD and 5000 BC positions of \PSRJ.

From {\em XMM-Newton} X-ray observations of 3C~58, Gotthelf et
al.\ \citep[2007; see also][]{Bocchino+2001} found an approximately
circular region with a softer X-ray spectrum, which they interpreted
as a thermal X-ray emission from the supernova shell.  They noted,
however, that the centre of this shell was displaced from the pulsar
position at \Ra{02}{05}{33}{97}, \dec{63}{49}{50}{0} (J2000.0).  We
plot this position also in Figure~\ref{f3c58img}.  Neither the present
distribution of synchrotron emission or thermal filaments suggests any
particular explosion centre, and although \citet{Fesen+2008}
determined proper motions for various optical features, the precision
is too low to accurately identify an expansion centre.

Our measured proper motion does not place \PSRJ\ near the centre of
the region of softer X-ray emission for either possible explosion
epoch.  For an explosion epoch of 5000~BC, however, the extrapolated
position of \PSRJ\ is closer to the geometrical centre of 3C~58.

If the area of softer X-ray emission identified by
\citet{GotthelfHN2007} were thermal X-ray emission associated with the
forward shock of the supernova, then one would probably expect its
centre to be near the location of the explosion and of \PSRJ's
birth. This does not seem to be the case.  Furthermore, the diameter
of the region of softer X-ray emission is only $\sim$5.6~pc, which is
notably smaller than the PWN, which has an E-W extent of $\sim$8.5~pc.
The PWN, however, is expected to still be confined by the ejecta and
thus be inside the forward shock, which suggests that the forward
shock is considerably larger than the region of softer X-ray emission.
Therefore, both because the region of softer X-ray emission is not
centred on the location of the explosion, and because it is small
compared to the PWN, we think it is unlikely that the softer X-ray
emission is associated with the supernova forward shock.

\section{Summary and Conclusion}

\begin{trivlist}

\item{1.} We obtained VLBI observations of \PSRJ, the pulsar in 3C~58\@.
We employed a novel technique to obtain VLBI observations of faint
pulsars, where we used the GBT simultaneously for pulsar timing
observations as well as an element of the VLBI array.  The derived
pulsar timing information was then used to gate the VLBI correlator
increasing the signal-to-noise of the pulsar VLBI.  This technique can
be used to advantage for other young pulsars which have high timing
noise.

\item{2.} We determined an accurate position for \PSRJ\ of
  \Ra{02}{05}{37}{92} \dec{64}{49}{41}{3}, which is 1\farcs5 different
  than the previously accepted one, which was based on an X-ray image.
  Re-examination of the X-ray data, however, reveal an error in the
  original reported position; the newly-determined X-ray position
  reported here is consistent with the VLBI position.  Furthermore,
  this position is coincident with an optical source identified by
  \citet{ShearerN2008}.

\item{3.} We determined the proper motion of \PSRJ. After correction
  for Galactic rotation, we found a proper motion of $2.3 \pm
  0.3$~\masyr, corresponding to tangential velocity of $(35 \pm 6)(D /
  3.2\,{\rm kpc})$~\kms\ at p.a. $-38$\arcdeg.  This low speed puts
  \PSRJ\ amongst the slowest few percent of young pulsars.

\item{4.} We estimated \PSRJ's position at birth.  If it is the
remnant of a supernova in 1181 AD, then its position at birth is only
1\farcs 9 different than at present.  If as seems more likely, the age
of \PSRJ\ and 3C~58 is several thousand years, then its position at
birth was near the midpoint of the presently visible nebula, although
somewhat displaced from the present radio brightness centre which is
close to the present location of \PSRJ.

\end{trivlist}

\section*{Acknowledgements}

The research at York University was supported by the National Sciences
and Engineering Research Council of Canada.  POS acknowledges partial
support from NASA contract NAS8-03060.
The National Radio Astronomy Observatory, NRAO, is a facility of the
National Science Foundation operated under cooperative agreement by
Associated Universities, Inc. The telescope at Effelsberg is operated
by the Max-Planck-Institut f\"{u}r Radioastronomie in Bonn, Germany.
This publication makes use of data products from the Two Micron All
Sky Survey, which is a joint project of the University of
Massachusetts and the Infrared Processing and Analysis
Center/California Institute of Technology, funded by the National
Aeronautics and Space Administration and the National Science
Foundation. We have made use of NASA's Astrophysics Data System
Bibliographic Services.

\bibliographystyle{mn2e}

\bibliography{mybib1,Scott_R.bib}

\end{document}